# Overcoming the hurdle of legal expertise: A reusable model for smartwatch privacy policies

Constantin Buschhaus [a], Arvid Butting [a], Judith Michael [a],*, Verena Nitsch [b], Sebastian Pütz [b], Bernhard Rumpe [a], Carolin Stellmacher [c], Sabine Theis [b]

[a] *Software Engineering, RWTH Aachen University, Aachen, Germany*
[b] *Institute of Industrial Engineering and Ergonomics, RWTH Aachen University, Aachen, Germany*
[c] *University of Bremen, Bremen, Germany*

## ARTICLE INFO

## ABSTRACT

Regulations for privacy protection aim to protect individuals from the unauthorized storage, processing, and transfer of their personal data but oftentimes fail in providing helpful support for understanding these regulations. To better communicate privacy policies for smartwatches, we need an in-depth understanding of their concepts and provide better ways to enable developers to integrate them when engineering systems. Up to now, no conceptual model exists covering privacy statements from different smartwatch manufacturers that is reusable for developers. This paper introduces such a conceptual model for privacy policies of smartwatches and shows its use in a model-driven software engineering approach to create a platform for data visualization of wearable privacy policies from different smartwatch manufacturers. We have analyzed the privacy policies of various manufacturers and extracted the relevant concepts. Moreover, we have checked the model with lawyers for its correctness, instantiated it with concrete data, and used it in a model-driven software engineering approach to create a platform for data visualization. This reusable privacy policy model can enable developers to easily represent privacy policies in their systems. This provides a foundation for more structured and understandable privacy policies which, in the long run, can increase the data sovereignty of application users.

## 1. Introduction

*Motivation.* Individuals are increasingly sharing personal data through various digital channels. This personal data includes sensitive information such as personally identifiable information and medical or financial data. While collecting and processing such data can benefit individuals and organizations significantly, it also creates substantial privacy concerns. Consequently, privacy protection regulations, such as the General Data Protection Regulation (GDPR) [1], aim to protect individuals from unauthorized data storage, processing, and transfer. In practice, however, the complexity of these regulations often makes it difficult for individuals with diverse educational backgrounds to fully grasp their meaning and act upon them in an informed manner.

To address this issue, it is necessary to provide privacy policies in a structured way. Providing privacy-related information for end users within software applications requires developers to extract their meaning on a conceptual level. Here, research has shown that it is not easy for developers to understand privacy concepts [2], that their perception of privacy is strongly influenced by security and mainly limited to third-party threats outside an organization [3]. Thus, there is a need for a conceptual model (in the following also called privacy policy model) that captures relevant concepts from privacy policies and can be easily integrated into applications by developers.

While providing privacy policies in a structured way is important for all domains where a privacy policy is needed, applications containing health data have a special need to be better understood. The GDPR recognizes data concerning health as a special category of data and requires specific safeguards for personal health data [4]. Smartwatches or other types of wearables collect health-related data that can be used for leisure, work, as well as medical applications, and they gain more and more popularity. Within 2024, 21.7% of the adult population has a smartwatch [5]. Thus, including better explanations of privacy policies in applications would be helpful for a large user base.

This paper focuses on integrating privacy policies into applications developed using Model-Driven Software Engineering (MDSE) methods [6] as an example for easy integration. The privacy policy model can be used as input for code generators and is then reflected in the application code for further processing. Integrating a conceptual model of privacy policies in the MDSE process ensures that all relevant concepts are covered.

Existing work focuses either solely on the GDPR without considering privacy policies of smartwatches, e.g., for checking GDPR compliance [7], checking for completeness [8], for verifying the correct implementation of security countermeasures [9], or validation of the technical design [10]. Other approaches focus on empowering end users to define their privacy preferences [11–13] but without a specific focus on the legal information companies provide. A third group of research topics focuses on the visualization of privacy policies [14] but lacks concepts for the privacy policies of smartwatches.

The leading *research question* of this paper is thus how to conceptualize privacy policies of smartwatches to be usable in MDSE approaches. With our solutions, we want to support the engineering of applications that aim to improve the understandability of privacy policies and data processing of smartwatches. This includes both, data processing by third parties and data shared between user-owned devices where users cannot be sure about what is exactly exchanged from a technical perspective.

The results presented in this article were developed as one part of the results of the InviDas project.[1] Within InviDas, we aimed to increase the digital sovereignty of smartwatch users by improving the understandability of privacy policies. The way privacy policies are formulated and presented up to now (purely textual, using specific legal terms) makes it hard for users to understand how their own data is processed. InviDas aimed at enabling users and future users of smartwatches to understand the main concepts within privacy policies as well as the general data processing happening with and without the users' active participation. The generated application should be able to show privacy policy concepts ergonomically and explain a privacy policy in detail. Additionally, the InviDas application allows users to compare different privacy policies. This article focuses on the developed conceptual model and its use for generating the main application.

*Contribution.* To support the building of applications that improve the understandability of privacy policies and data processing of smartwatches, we introduce a conceptual model for privacy policies and related information for smartwatches and demonstrate its use in a MDSE approach to create an application that can be used to explain and visualize privacy policies ergonomically. For being able to create the conceptual model, we have analyzed the privacy policies of various wearable manufacturers and extracted the relevant concepts. For the presentation of the results within this article, we follow the suggestion by Mayr and Thalheim [15]: We present the conceptual model as a class diagram, define the notion space, and show concrete examples from our evaluation of actual privacy policies of smartwatches. Moreover, we have checked the model with lawyers for legal compliance. The model was instantiated in two instances with concrete data from Garmin's privacy policy for their smart wearable services and with FitBit's privacy policy. Please note that some privacy policies, e.g., from Garmin or Fitbit, do not differentiate between data from smart wearables and smartwatches, as they have one privacy policy for all their smart wearable products. We have employed the model in a MDSE approach, creating a platform for data visualization of wearable privacy policies from different smartwatch manufacturers. This model makes it possible to create systems that automatically include relevant concepts for privacy policies.

*The paper is structured as follows:* Section 2 discusses the term *privacy*, related work, and the generator framework used for representing the conceptual model in a software system. Section 4 describes how we have developed the conceptual model. Section 5 presents the conceptual model and notion space. Section 6 shows the application of our approach for generating the InviDas platform and the visualized privacy policies. Section 7 discusses our approach, and the last section concludes.

## 2. Background and related work

To lay the groundwork for our approach, we begin with the fundamentals of the term privacy, then go into methods for modeling privacy, and explain the technology used to validate our approach using MDSE methods.

### 2.1. Privacy

The term *privacy* refers to the non-public area in which a person exercises their right to free development of personality without external disturbances [16]. In the digital realm, this concept is also known as the right to informational self-determination. Privacy is considered a basic human right and related to the ability to decide what information about a person goes where [16]. In Europe, the General Data Protection Regulation (GDPR) mandates that organizations must consider privacy throughout the entire development process [1]. Consequently, privacy considerations in software and system engineering are crucial to ensuring it in the digital context, e.g., via using privacy-friendly design methods for systems and services [17]. Software security serves as the essential foundation for ensuring data privacy by preserving the confidentiality, integrity, and availability of information, as well as supporting authenticity, accountability, non-repudiation, and reliability [18,19].

---

[1] funded 2020–2023 by the German Federal Ministry of Education and Research (BMBF) https://www.interaktive-technologien.de/projekte/invidas.

Privacy-by-design principles, such as the integration of privacy requirements as early as possible in the development process, aim to improve the overall privacy friendliness of IT systems [20–22], from the initial planning to the final testing and deployment [23]. Privacy-by-design principles encompass technical, organizational measures to prevent data breaches [24], to provide clear and transparent information about data collection and use [25], and to give users the uttermost control over their personal data [26]. Special attention has been given to the privacy of ubiquitous and wearable systems [27–29] due to their collection of sensitive data and health information. But to better embed privacy in the development of such systems, companies need to value privacy-by-design [30], technical implementations must be simplified, and privacy frameworks have to be applied in different contexts [26,31].

Although privacy-by-design approaches have been widely discussed, they effectively illustrate how to enhance developers' privacy awareness and facilitate the implementation of privacy-enhancing technologies (PET) [32]. Even more than general design recommendations, concrete models provide a fundamental basis for software development, as they can represent objects with values and behavior [33], layers to store business logic and processes [34] to build and understand a software [35].

*2.2. Privacy and modeling*

Model-driven approaches to engineering privacy-preserving software have primarily focused on Data Flow Diagrams (DFD) for risk classification [36–38], tracking the movement of personal information within IT systems [39], or privacy checkpoints within the system architecture [12,40]. However, these methods often fall short of ensuring legal compliance throughout a system's lifecycle.

Although privacy has been relatively underexplored in MDSE [41], it can be used to add privacy concepts already in the software code. Michael et al. [12] have developed an MDSE approach for systems using policy-based access control where they model privacy-related concepts, e.g., user roles such as data providers and data controllers, privacy policy rules, and data usage purposes to decide in a generated application which data should be provided to data consumers on request. This approach lacks to explicitly consider GDPR concepts. Early approaches have addressed areas such as privacy risk identification [42], privacy-oriented data management of intelligent assistants [43], privacy-driven process mining system design [44], or model-based privacy-related continuous risk control [45]. Additionally, the Model-Driven Service Engineering Architecture method has been used to enrich business processes with technical security information [46], while function-based modeling has supported the generation of health monitoring applications [47]. Such conceptual models for Software Privacy-Enhancing Technologies (PETs) vary significantly in their comprehensibility and technical detail based on the application focus.

Torre et al. [8] developed a technical-oriented model by analyzing a total of 234 privacy policies to characterize the information content envisaged by GDPR, which was then used to assess compliance with GDPR using artificial intelligence similar to machine-readable privacy certificates for banking services [48]. However, while the model by Torre et al. [8] aids in GDPR compliance through detailed traceability and validation, a user-centered model for making privacy policies more understandable, especially for smart wearables, is lacking. Another conceptual model mapped GDPR requirements to security controls set by the National Institute of Standards and Technology (NIST) to ensure correct implementation of security measures [9]. Models have also been developed to secure information flow in complex systems, adhering to privacy-by-design principles [49], and to validate technical designs concerning European privacy regulations [10]. Krasnashchok et al. [50] presented an ontology for describing GDPR concepts. However, they have no specific focus on smartwatches.

The only comprehensive Unified Modeling Language (UML) model of the GDPR includes traceability from its classes to the regulation, a glossary, and a rule description [7]. This model is less concerned with the comprehensibility of the language since its primary purpose is to automate compliance checking of privacy policies against the GDPR and they have no specific focus on smartwatches. In contrast, the TILT language model [13] focuses on increasing transparency and understanding for end-users by making GDPR-related declarations more accessible and comprehensible. Other models empower end-users by enabling them to define their privacy preferences and adjust the applied privacy policies accordingly [11], or by analyzing privacy settings in social media applications [51].

Breaux, Vail and Antón [52] developed methods for semantic parameterization to extract rights and obligations from legal texts like the Health Insurance Portability and Accountability Act (HIPAA) privacy rule, to clarify ambiguities, and to align software systems with essential privacy and security requirements. However, their approach primarily addresses the early stages of software development and overlooks ongoing compliance challenges.

Sartoli, Ghanavati, and Namin [53] introduced a framework for continuous compliance management in dynamic environments, like cloud computing, where legal requirements may change. This framework integrates legal requirements into the Goal-oriented Requirements Language (GRL) model, enabling continuous monitoring and adaptation. Our work extends this by integrating compliance monitoring into both static and dynamic systems. Alshugran and Dichter [54] proposed methods for extracting and modeling of HIPAA privacy requirements within the software lifecycle to reduce later costs and complexities. However, this approach lacks comprehensive traceability and management of legal obligations throughout the entire product lifecycle. Breaux, Antón, and Spafford [55] also developed frameworks to manage compliance and accountability by distributing legal obligations across business units and ensuring traceability, but these frameworks primarily focus on the organizational level rather than integrating compliance into the software's technical layers. Ghanavati, Amyot, and Rifaut extended the GRL framework to model legal requirements, allowing systematic compliance analysis but focusing more on static compliance, without fully addressing continuous compliance in dynamic environments [56]. Rashidi-Tabrizi, Mussbacher, and Amyot [57] developed the Measured Compliance Profile for GRL, which formalizes legal texts for compliance analysis and evaluates software systems using real-world measurements. While useful for compliance verification, it does not integrate with business processes or support continuous adaptation to changing legal requirements. Ghanavati, Amyot, and Peyton further explored aligning business strategies with legal regulations, particularly in healthcare [58]. Although this work highlights the importance of this alignment, our approach goes further by integrating compliance into the software development process, ensuring ongoing compliance as systems evolve.

In addition to legal considerations of data handling in general, the domain-specific use cases and their data structures influence the conceptual model. Only Torre et al. [8] refer to existing privacy statements. Conceptual models for smart wearables include a conceptual model leveraging blockchain to improve the security, privacy, interoperability, and data integrity of FDA-approved medical wearable devices [59]. There is a clear need for conceptual models based on privacy policies from different smartwatch manufacturers to support developers in the engineering process of related applications. The unique characteristics of smartwatch privacy, including the collection of highly sensitive and continuous data, integration with other devices, regulatory compliance, and the need for user-centric privacy management, necessitate the design of a specific conceptual model to address these challenges effectively.

*Privacy Policy Taxonomies.* Morel and Pardo [60] conducted a meta-study on variants of privacy policy depiction forms regarding their benefits and limitations. Their categorization consists of three forms ("facets"): natural language, machine-readable and graphical. Their assessment of the benefits and limitations was based (among other things) on an adapted privacy policy taxonomy first introduced by Wilson et al. [61]. We propose a meta-model for privacy policies as Class Diagram (CD), thus our model and its instantiations (UML Object Diagrams) are machine-readable and we can compare our model to the inspected machine-readable language solutions. Unlike the privacy policy languages analyzed by Morel and Pardo, which are designed for automatic enforcement, auditing, and similar purposes, we primarily aim to use our conceptual model within an MDSE approach. Furthermore, our model is specifically meant to provide support for visualizations and has already proved to be useful in this regard (InviDas application, see Section 6). CDs are better qualified for providing the input of an MDSE approach than (domain-specific) languages as we can reuse existing tool infrastructures such as parsers or validators and do not have to develop them newly for a domain-specific language. Furthermore, the infrastructure generators such as MontiGem that take CDs as input can help to automatically create a large part of it. Aside from these advantages, in contrast to the inspected languages from Morel and Pardo [60], our model supports the majority of concepts of their taxonomy (see Section 7.1) while also having an existing implementation. According to Morel and Pardo, no machine-readable language they inspected provided the means to include a lawful basis for data processing or a way to indicate how a privacy policy may be changed. However, the languages may have evolved since their analysis and may provide such concepts now, either directly or included in other concepts such as purposes.

## 2.3. MDSE with the MontiGem generator

Besides using conceptual models for understanding the domain in the analysis phase, they can be reused in MDSE processes to synthesize code. Within this work, we use the generator framework MontiGem [62] to create an information system based on our conceptual model. MontiGem is based on the MontiCore language workbench and code generation framework [63]. As input models, MontiGem can handle MontiCore Domain-Specific Languages (DSLs) [64], e.g., a UML Class Diagram describing the relevant domain concepts, models for Graphical User Interfaces (GUIs) [65] and data aggregation within these GUI models as well as models for data input validation in OCL. These models are transformed into source code [66], namely to Java for the backend and TypeScript and HTML within the Angular framework for the frontend. The result is a running information system including a relational database, and the backend and frontend of the application. Mechanisms to access the concepts from the conceptual model in the database are provided in the backend and we generate the transport infrastructure to send information between the backend and frontend of the application. The generated platform can be extended by hand-written code [67] to implement further functionalities we are not able to generate automatically, e.g., business logic such as domain-specific calculations. Up to now, MontiGem has shown its applicability in a series of research and industry projects for different application domains, e.g., for financial and staff management [68], to support the model-based engineering of wind turbines [69], energy management systems [70], IoT app stores [71], cockpits for self-adaptive digital twins [72], and in research projects on privacy-preserving information systems [12], process-aware information systems [73], low-code platforms [74] and assistive systems [75,76].

Within the InviDas project and the approach presented in this article, we use the generator framework MontiGem to create a platform for improving the understandability of privacy policies of smartwatches. MontiGem takes the conceptual model describing privacy policies (see Section 5) as one of its inputs and creates the data structure, backend, communication infrastructure between frontend and backend, and parts of the frontend of an application to explain and ergonomically visualize the privacy policies (see Section 6). More complex visualizations were added as hand-written code in the InviDas application while still referring to the same data structure and using the generated communication interfaces.

## 3. Our approach for reusable privacy policy concepts

To support developers, we propose a method for reusing privacy policy concepts for smartwatches within software engineering processes. First, we show the current challenges for developing applications that both, require and process privacy policies for smartwatches. Our applications in focus use data from smartwatches. This could be, e.g., applications for representing smart watches and monitoring data for dementia patients, depression detection, or applications comparing running data from smart watches for competitions and online games. Using this smartwatch data requires the applications to handle privacy policies. This results in the requirement that end users should be enabled to understand the privacy policies provided so that they can give their informed consent. In the next step, we introduce modeling privacy policies as a solution for these challenges and describe its requirements.

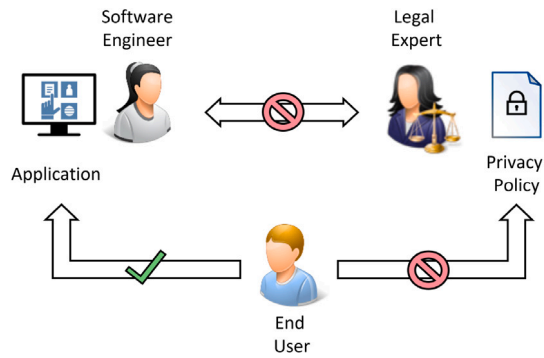

**Fig. 1.** There are three interest groups in the context of the PPM, that have difficulties to understand each other. The arrows indicate, that one interest group has to process information of the arrow's target. A red sign indicates problems in this processing.

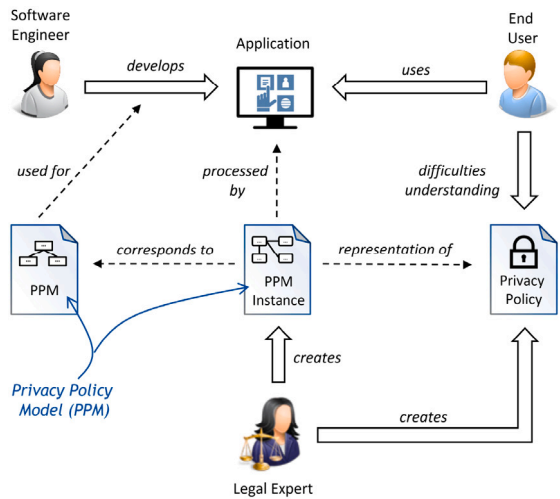

**Fig. 2.** Overview of the PPM usage by its different interest groups.

The applications in our focus have three main groups of stakeholders: software engineers, legal experts, and end users (Fig. 1). Their differences in expertise create challenges for the software development process and application use. Developing applications using smartwatch data requires cooperation between software engineers and experts from the legal domain including their terminology and expertise, while end users have difficulties comprehending textual privacy policies defined by legal experts [77].

To tackle these challenges, we propose the use of a model for describing privacy policies for smartwatches (PPM). A PPM describes relevant concepts of privacy policies of smartwatches from the legal perspective. Privacy policies of smartwatches cannot be directly used as input for the development process of software applications because they are legal text written in natural language. Different policies can be very similar in meaning while being syntactically entirely different (see a comparison of privacy policies in Section 4). The PPM solves this problem by specifying the structure and content of its instances, where each instance is itself a model of a textual privacy policy.

Fig. 2 shows how a PPM improves the current situation (c.f. Fig. 1 for current difficulties in understanding) for the three groups of stakeholders. In the following, each group is examined in more detail, to explain its interactions with the PPM.

The *end users* (top right corner in Fig. 2) of the applications have no legal expertise and they want to know about the processing of their data according to a privacy policy. This group does not have contact with the PPM itself (only via instances, which were already processed by an application), but it is the overall target group of the application.

The second group are *legal experts* in charge of creating and updating privacy policies, e.g. data protection officers, for companies that handle personal data. Ultimately, someone has to take a privacy policy to create an instance of the PPM, so that the instance can be used by applications based on the PPM. The company or its personnel are best qualified to create an instance of the PPM based on their textual privacy policy because, in addition to the privacy policy, they have access to background knowledge about technical aspects and internal data processing. Therefore, they do not have to interpret the textual privacy policy and may add further information as intended by the PPM. The creates-arrow from *legal expert* to PPM instance in Fig. 2 is a simplification. In reality, legal experts will probably need support for the creation of PPM instances. One exemplary method to provide such support are online forms such as Fig. 11 that can be developed with the help of the PPM.

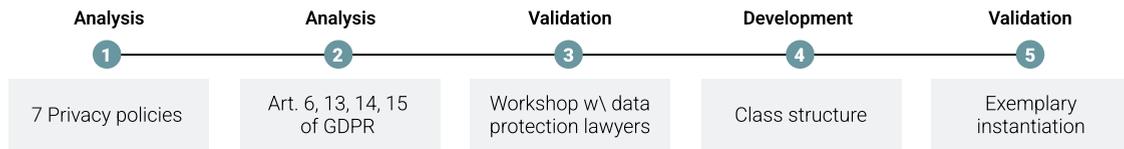

**Fig. 3.** Five-step process that informed the conceptual model for smartwatch privacy policies.

Lastly, the third group is the group of *software developers* of the application using the PPM. Once developed, the PPM can be reused by software engineers in the software engineering process of an application (see Fig. 2). There are two ways to use the PPM: (1) The PPM is used in model-based approaches serving as information for requirement analysis and about concepts that have to be considered for developing an application processing privacy policies and generally to cope with the complexity of privacy policies. (2) The PPM is used in MDSE approaches as input for a code generator creating parts of the application in an automated way Fig. 10. Once the PPM was created and validated by legal experts, it reduces the involvement of legal experts in the development phase of applications reusing the PPM, because the PPM already captures their domain-specific knowledge about concepts and structures of privacy policies.

However, during the runtime of the application, legal experts have to be involved with the creation of instances of the PPM. Depending on the use case, an application may process multiple instances of the PPM at the same time, e.g., when a new version of a PPM instance (i.e. a new version of the privacy policy) is added, or when an existing version is updated, or when the application processes privacy policies from different vendor e.g. to make them comparable. Transferring privacy policies into instances of a common model reduces the legal domain from a space of many uncountable possibilities of unstructured data into a structured space. The main task is evaluating which concept of the model fits best to concrete excerpts of a privacy policy. A desirable side effect of this is that legal experts are forced to simplify and align their policies during the transformation to fit them into the model. This increases the policies' comparability since they all have to comply with the same uniform structure.

A privacy policy model that enables software development for our applications in focus needs to fulfill the following requirements:

R1  It must specify the structure and the content of privacy policies in such a way that they can be automatically processed by a computer.
R2  It must cover important legal concepts existing in the legal requirements for privacy policies.
R3  It must cover the large variety of existing privacy policies from different manufacturers of smartwatches.

These requirements have to be taken into account in both, the development process of the PPM as well as the development process of the application. Additionally, to facilitate the incorporation of the conceptual model in a software engineering process, there is a technical incentive to leverage long-standing research and existing tooling support (such as MontiGem Section 2) for agile software development based on the Unified Modeling Language [67]. As a result, the model should be defined as a UML/P CD.

## 4. Five-step process to derive the conceptual model

We applied a five-step process (see Fig. 3) to identify and collect concepts from privacy policies and to better understand the characteristics of data processing that a conceptual model must consider. To do so, we analyzed the existing privacy policies of smartwatch manufacturers (Step 1) as well as the GDPR (Step 2) to extract all legal requirements. The derived list was then verified with lawyers (Step 3). Based on our findings, we created a class diagram (Step 4), which we validated through two exemplary instantiations (Step 5). The result is our conceptual model presented in Section 5. Details for each step are described in the following.

In step 1, we analyzed the privacy policies of seven of the major smartwatch manufacturers, namely Apple, Fitbit, Fossil, Garmin, Huawei, Samsung, and Xiaomi. The privacy policies were examined for similarities and differences in information content to extract general concepts as a basis for the model structure. To do so, we combined a top-down and a bottom-up coding approach, where we used concepts from the GDPR as starting points and extended them with further types of information provided in the privacy policies. This approach resulted in a total of ten information categories: (1) data type, (2) data processing, (3) data owner, (4) data source, (5) data receiver, (6) processing purpose, (7) processing decision, (8) data protection, (9) data storage period, and (10) legal basis. To give an overview of the obtained results, Table 1 provides exemplary instances for the identified information categories and the total number of unique instances across the seven privacy policies. The number is often presented as a range since the differences in wording between privacy policies made it difficult to assess whether the used terms were synonymous or not.

In step 2, we have identified the legal requirements imposed by the GDPR and extended the findings from the previous step. We focused mainly on Article 6 ("Lawfulness of processing"), Article 13 ("Information to be provided where personal data are collected from the data subject"), Article 14 ("Information to be provided where personal data have not been obtained from the data subject") and Article 15 ("Right of access by the data subject"). These articles explicitly state the rights of users and the required or optional information about personal data that manufacturers need to disclose through privacy policies. By combining steps 1 and 2 for the

**Table 1**

Information categories identified in the analysis of privacy policies of smartwatch manufacturers (N = Number of different values per category across the seven privacy policies; the same value can occur in multiple policies but is only counted once).

| Information category | N | Exemplary instances |
| --- | --- | --- |
| Data type | 178–207 | personal data, health data, location data, device data, name, phone number, heart rate |
| Data processing | 15–23 | collection, storage, usage, merging, processing, review, transfer, disclosure, deletion |
| Data owner | 4 | user, other user, shared contact, third party |
| Data source | 7–9 | provided by the user, collected during usage, shared by other users, collected by third party |
| Data receiver | 14–22 | manufacturer, subsidiary, service provider, mobile carrier, other user, investor, authorities |
| Processing purpose | 70–99 | providing service, developing service, sending notification, customer support, marketing |
| Processing decision | 10–15 | voluntary, mandatory, give consent, withdraw consent, delete data, when a function is used |
| Data protection | 14–16 | data encryption, physical protection, technical protection, administrative protection |
| Data storage period | 7–10 | active account, required for service, not actively deleted, legal retention period, X month |
| Legal basis | 5 | consent, performance of a contract, compliance with a legal obligation, vital interests |

analysis, our developed model incorporates both explicit legal requirements and existing practices for providing privacy-related information to end-users that are implemented in the industry.

Findings from the analysis were then evaluated, in Step 3, during a workshop with lawyers specialized in data protection and information technology law. The discussion reflected on the validity and comprehensiveness of our set of privacy information, clarifications of specific legal meanings, discrepancies between the analysis findings from the privacy policy and GDPR, and open questions. Having identified and evaluated the information basis for the conceptual model, we developed a corresponding class structure in step 4, including mapping all identified privacy information categories to class attributes and specifying class associations to model their interdependencies. The resulting class diagram was the first full representation of the conceptual model.

Finally, in step 5, we validated the class diagram through an instantiation with concrete data. This was done through a systematic and extensive examination of Garmin's privacy policy, including 4 researchers of different domains (Software Engineering, Human–Computer Interaction, Ergonomics, Psychology). By doing so, we iteratively confirmed and adjusted the conceptual model to hold the relevant privacy information of smartwatch privacy policies. For aspects where we were unsure, we involved legal experts who checked our model and provided additional context as to why certain aspects were worded as they were. For validation purposes, we repeated the process of mapping a concrete privacy policy to the conceptual model, namely the Fitbit smart wearable privacy policy.

## 5. A conceptual model for smartwatch privacy policies

Conceptual models are created for various purposes. The major purpose for creating the conceptual model for privacy policies of smartwatches presented in this chapter is to structure textual privacy policies. This structure is used as a basis for designing applications that handle documents that often comprise lengthy and complex formulations. Hence, it is not in the focus of this conceptual model to be able to check individual instances of the model against any legislation. In other words, we do not guarantee that all instances of the model are legally compliant or even legally binding, e.g., with the GDPR.

The conceptual model for privacy policies of smartwatches, visualized as a UML Class Diagram, is depicted in different figures throughout this section that each show a part of the whole conceptual model. The remainder of this section describes the most relevant concepts of the conceptual model and an excerpt of their attributes in more detail. It illustrates them by concrete examples based on the Garmin privacy policy for their smart wearable services.[2] The privacy policy excerpts that each example is based on are provided in Table 2 and referenced in each example. The detailed conceptual model including attributes and enumerations and the two instances as object diagrams can be accessed at Zenodo [78]. To provide a usable conceptual model for law systems where the GDPR or equivalent laws do not hold, the model allows fuzziness to some extent in parts that are heavily derived from the GDPR, e.g., a not applicable option in enumerations based on GDPR laws. Furthermore, the privacy policy model allows adding information that cannot be captured in a structured way in various places, e.g. in the form of description attributes.

### 5.1. Basic information in privacy policies

The conceptual model (see Fig. 4) realizes the basic information of a privacy policy as direct attributes of the class `PrivacyPolicy`. Each privacy policy in the conceptual model has a *name*, a *full text*, and the *name of the manufacturer*. The official textual version of the (*original*) privacy policy is usually available online and the conceptual model refers to this source via a URL. Furthermore, each privacy policy has an *effective date*. This must not be confused with the *date of the last change* in the conceptual model instance. The *date of creation* and *version* attributes give insights into the history of the privacy policy. According to the GDPR, a person is required to be of a certain *minimum user age* to be able to give their consent to a privacy policy. This minimum age can vary depending on each country's law. Manufacturers handle updates to privacy policies differently. To reflect this in the conceptual model, each privacy policy has an attribute with a description of the *update policy*.

---

[2] https://www.garmin.com/en-GB/privacy/connect/policy/

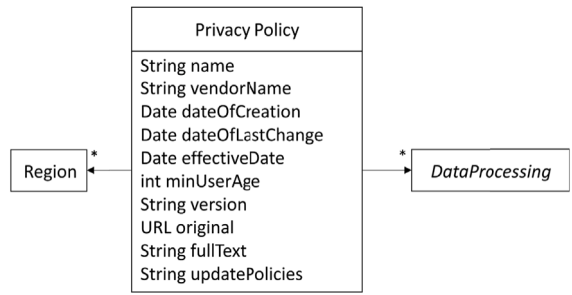

**Fig. 4.** Privacy policy concept as main starting point (part of the conceptual model of privacy policies).

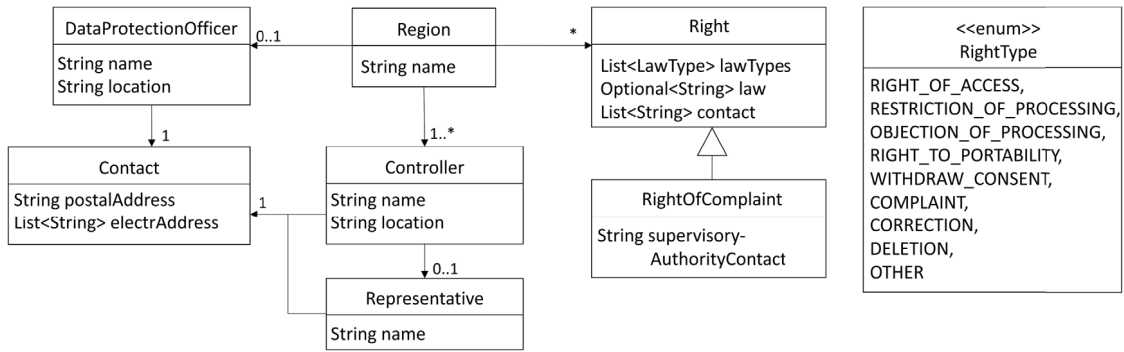

**Fig. 5.** Regions and contact persons for privacy policies (part of the conceptual model of privacy policies).

**Example.** An instance of the `PrivacyPolicy` class from the Garmin privacy policy could be as follows. It is named "Privacy Policy for Garmin Connect and Compatible Garmin Devices", the vendor name is "Garmin Ltd"., and the date of the last change could be August 31, 2022. The effective date could be June 05, 2020. The minimum user age for the Garmin privacy policy is 16. It is the third version of this privacy policy and changes to the last version are contained in the update policies attribute. The URL to the original refers to the Garmin website. (For the concrete privacy policy excerpt see Privacy Policy in Table 2.)

*5.2. Regions*

Privacy policies contain explanations of compliance with regulations and information about the data processing specific to a certain `Region`. A single privacy policy usually describes compliance with regulations for multiple regions at once. In the latter case, each region may have individual contact persons. Each manufacturer can define their regions, a typical example is the European Union as one region as they have a common ground of laws regarding privacy policies defined in the GDPR. In the conceptual model (see Fig. 5), each region has a *name* that identifies the region. Moreover, each region must have at least one person that acts as `Controller` for this region and that must be contactable via publicly accessible `Contact` data, such as e-mail or postal addresses. Optionally, each controller can have a `Representative` person with individual contact data. Beyond the controllers, there may be an optional `DataProtectionOfficer` for each region also available via specific contact data.

**Example.** The Garmin privacy policy applies to different regions. The Garmin privacy policy summarizes the European Economic Union, the U.K., and Switzerland into one `region`. The `Controller` for this region is Garmin Würzburg GmbH and it is located in Germany. Its postal address is Beethovenstr. 1a, 97080 Würzburg, Germany. The `data protection officer` has the same name and location and its contact mail is "euprivacy@garmin.com". (For the concrete privacy policy excerpt see Region from Table 2.)

In each region, there may be legal regulations regarding data processing. Privacy policies usually describe user/customer options to make use of such regulations in terms of certain `Rights`. For the context of privacy policies, we distinguish several kinds of rights (*right types*), which were identified in our analysis (see Section 4). These right types relate to data deletion, restriction of data processing, right of access, right of correction, objection to data processing, right to withdraw one's consent, and the right to access one's data. Since the assignment to exactly one right type may not always be possible in different regions, the right type *other* was added. Each right in the conceptual model has a (set of) contact address(es) and an optional descriptive text. A special type of right is the *right of complaint*, which is exercised with a supervisory authority and it therefore needs to have a contact with the authority. `Right` has an optional attribute *law* that can be used to add more information about the law from which rights for data providers are derived from. This is especially relevant for the right type *other*, which can be selected if none of the other *right type* enumeration options is applicable.

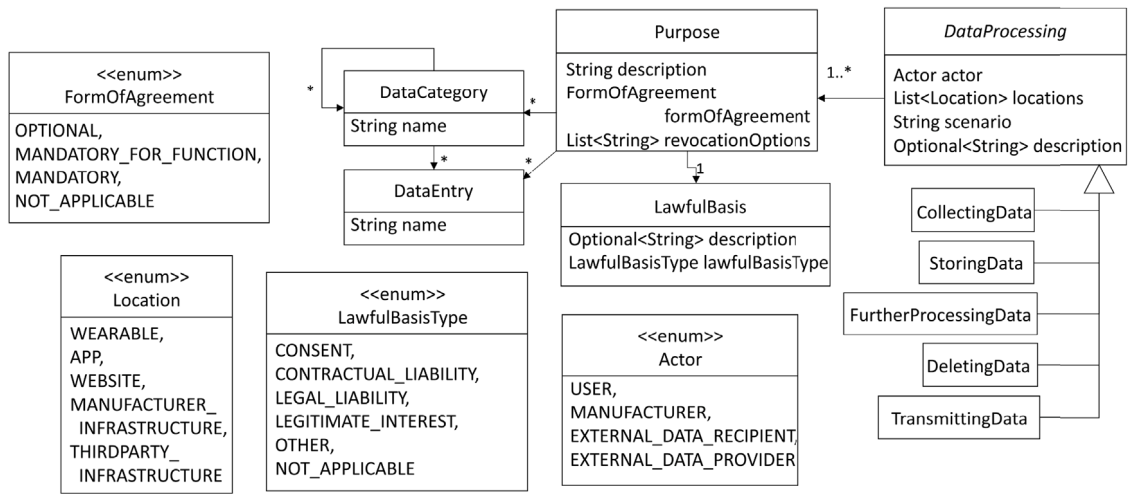

**Fig. 6.** Concepts for data processing and the purposes of the processing (part of the conceptual model of privacy policies).

**Example.** In the Garmin connect privacy policy, the user/customer's rights for the following right types correction, deletion, and the right to portability can be exercised in the same way, so they are represented by one right. The privacy policy does not contain the law text or a further description of these laws. The Garmin privacy policy states that the user can make use of his rights, which are derived from the aforementioned right types, by visiting the Account Management Center, which acts as a form of contact. (For the concrete privacy policy excerpt see Rights in Table 2.)

*5.3. Processing data*

This subsection gives an overview of the abstract concept of data processing in our conceptual model and shows what all concrete processings have in common. Each form of processing data described in a privacy policy concerns certain kinds of data. For instance, collecting data from a user device may concern profile data, sensor data, a postal address, or a user name. However, the level of detail in which such data kinds are described can vary between different privacy policies and even between sections of a single privacy policy. Therefore, the conceptual model (see Fig. 6) distinguishes concepts for `DataEntry` and `DataCategory` that are both identifiable via *names*. A data category can contain one or more data entries. In the opposite direction, one data entry can belong to different categories. Moreover, our conceptual model allows data categories to be decomposed from other data categories. The distinction between data entry and data category in the model allows for treating them differently. Different manufacturers may consider different data entries to belong to certain data categories. Therefore, the model for data kinds is not fixed and each privacy policy defines its own data categories and data entries.

**Example.** An e-mail address (data entry) can belong to a data category "profile data" and at the same time, to other data categories such as "contact data", or "user data", which can all be categorized as "personal data". (For the concrete privacy policy excerpt see Data Entry in Table 2.)

*5.4. General information about processing data*

The conceptual model (see Fig. 6) distinguishes several forms of processing data that share common attributes. Hence, the concept `DataProcessing` is modeled as an abstract class and the concrete forms are modeled as concrete classes that extend the abstract class. The abstract class contains commonalities in terms of attributes and associations with other concepts. Each section of a privacy policy that describes data processing has to contain an *actor* and a *location*. We identified four types of actors, that may be involved in different kinds of data processing: the user themself, the manufacturer, an external data recipient, and an external data provider. The possible locations, where processing can happen, are the wearable, the app (usually used accompanying a smartwatch), the manufacturer's website (things like account creation or user support requests are often handled via the website), the manufacturer's infrastructure, and third-party infrastructure. The conceptual model uses *scenario*-based descriptions of data processing to foster the understandability for users and logical grouping of data processing.

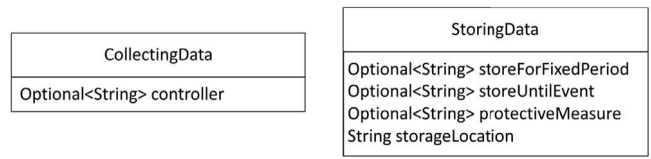

**Fig. 7.** Specializations of data processing: Collecting Data (left) and for Storing Data (right) (part of the conceptual model of privacy policies).

**Example.** Consider the scenario of a user contacting Garmin's product support. The user could contact Garmin e.g. by mail. To be able to answer the user, Garmin collects their e-mail address. Since the user's e-mail is collected, the user is considered to be the actor in this scenario. The location of data collection is Garmin's website or infrastructure depending on how the user contacts Garmin. (For the concrete privacy policy excerpt see General Data Processing in Table 2.)

The data processing in a specific scenario must have at least one (can have more) `Purpose`, which refers to the data entries or categories, that are processed for this specific purpose. Each purpose has a `LawfulBasis`. For the purposes, the conceptual model contains attributes, such as a purpose *description*, the *form of agreement*, and options/ further explanations for *revocation* of the agreement (if it is possible). There are three forms of agreement to a purpose. Firstly, there is an optional agreement, which can be rejected without any impact on the user experience. Then there is an agreement for specific functionality. These agreements can be rejected, but rejection entails, that the user cannot use some services/functionality anymore. Lastly, there is the mandatory agreement. This agreement can only be rejected by not using the wearable, at all. For the lawful basis, we distinguish between several `LawfulBasisTypes` and provide the possibility for an optional *description*. The different lawful basis types, that were identified in our analysis are consent, contractual liability, legal liability, and legitimate interest.

**Example.** Picking up the example from above of a user contacting Garmin's product support. The purpose of this data processing can be described by "This information is used to provide customers with product support". The form of agreement is mandatory for function with the function being product support. There are no options to revoke this processing since the data collection cannot be undone. The lawful basis is a legitimate interest, which can be explained in the description to provide product support for their customers. (For the concrete privacy policy excerpt see Purpose in Table 2.)

*5.5. Specific forms of processing data*

The conceptual model distinguishes between five different concepts for different forms of processing data, which are `CollectingData`, `StoringData`, `TransmittingData`, `DeletingData`, `FurtherDataProcessing`.

The latter is an aggregation of data processing forms, such as transformation, analysis, organization, alignment, combination, and others.

*5.5.1. Collecting data*

When `CollectingData` (left side of Fig. 7), the *actor* and the *location* of the data processing concept refer to the actor and location from which the data are collected. For instance, if the GPS position is recorded during an outdoor activity, the data collection may be triggered by the user and the location is the smartwatch itself (or a certain sensor of the smartwatch if the level of detail in the model allows distinguishing different locations within a smartwatch). Apart from the attributes for the general information about data processing, the concept of collecting data has an attribute describing a *controller*, which is responsible for the collection.

**Example.** A data collection happens in the scenario of adding a new device to the user's Garmin account. The data is collected from the user and it takes place in the Garmin app or on their website. The data collection does not have an extra controller. (For the concrete privacy policy excerpt see Collecting Data in Table 2.)

*5.5.2. Storing data*

When `StoringData` (right side of Fig. 7), the *actor* and the *location* of the data processing concept refer to the actor and location that initiate the data storage. Data can be stored either for a *fixed period* or *until an event* occurs. In the first case, the conceptual model has an attribute to indicate, e.g., that certain data is stored for 48 months. For the latter case, e.g., if a user deletes her/his user account, a short text describes the event in the conceptual model. Optionally, some privacy policies indicate certain forms of *data protection* for the storage, e.g., regarding the actual machine or the format in which the data is stored. Furthermore, the data storage indicates a *storage location* in the form of a country code.

**Example.** In the scenario of a customer uploading data to their Garmin account profile from Europe, Garmin is considered the actor, and Garmin's server infrastructure is considered to be the location. Garmin stores this data on servers located in the US, UK, or AU and they store it as long as the user account is considered to be active until the event of the account going inactive. There is no information given in the privacy policy about what security measures Garmin implements regarding this data storage. (For the concrete privacy policy excerpt see Storing Data in Table 2.)

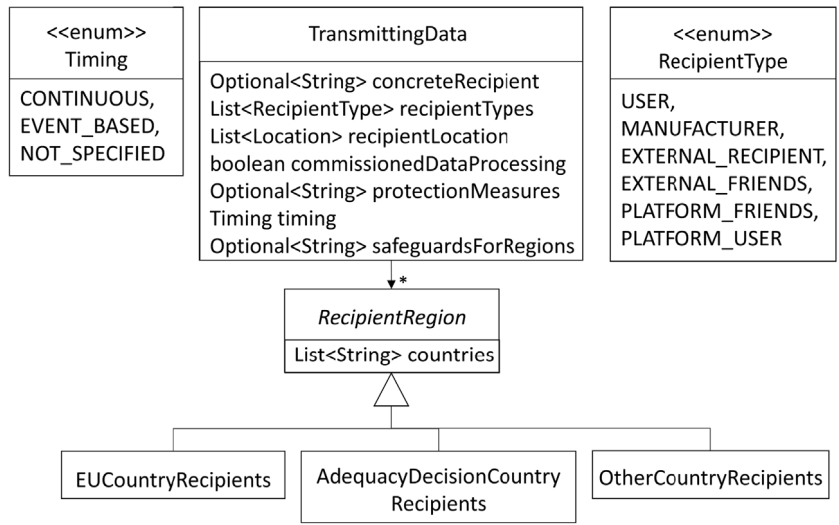

**Fig. 8.** Specialization of data processing: Transmitting Data (part of the conceptual model of privacy policies).

#### 5.5.3. Deleting data

The conceptual model does not foresee any information specific to data deletion. However, the general attributes of the concept of data processing apply.

**Example.** *Garmin deletes children's data, in the scenario of Garmin having the data unlawfully collected because the child was under the age of 16 in the EU. The actor would be the user in this context also including a legal guardian and the location can be the manufacturer's infrastructure.* (For the concrete privacy policy excerpt see Deleting Data in Table 2.)

#### 5.5.4. Transmitting data

From a legal viewpoint, data transmission is the most complex form of data processing. There are two different understandings of data transmission, one from the technical and the other one from the legal point of view. From the legal point of view, data transmission occurs when data is sent to another legal entity other than the one that initially collected the data. From a technical point of view, data transmission occurs, when data is sent from one location to a different one. The conceptual model does not commit to one interpretation and supports modeling both, as they are both reasonable in different contexts. The technical point of view is useful e.g. to disclose security-relevant information to the user. A manufacturer may collect data and only store it on the user's device or further process it on their server infrastructure. This would legally not be regarded as a data transmission, since the data is collected and processed by the manufacturer. However, it is important information for the user to evaluate their data security. They are in control of their own device but not of the manufacturer's server infrastructure or the communication between the device and the server.

`TransmittingData` in the conceptual model (see Fig. 8) uses the actor and location of the abstract class for the sender of the data transmission and introduces new attributes for the *actor* and *location* of the data recipient. Moreover, the recipient is classified based on a distinction into different *recipient types*, such as the user, the manufacturer, or a third party. The data transmission has a certain *timing*. There are three possible timings: continuous, event-based, and unspecified if the original privacy policy does not contain information about when the data is collected. The difference between continuous and event-based is that, in the case of continuous timing, the user has to actively revoke their consent to stop the data processing. Optionally, instances of the conceptual model may indicate *protection measures* for each data transmission.

If the data is transmitted from a legal point of view, more information must be indicated according to the GDPR. Whether it is a commissioned data processing i.e. the receiving third party has no freedom of decision on how to use the received data. And secondly, the specific reception areas have to be stated. For these, we distinguish recipient areas inside the European Union, locations with adequacy decisions, and other locations. For these other locations, there can exist guarantees that can be captured by the *safeguards for region* attribute [1, cf. Article 13 §1.f].

**Example.** *Smartwatch users of Garmin devices can create live tracking sessions of their devices and share this information with other users. This is a technical and legal data transmission, too, because other people than the original user get access to the user's location data. The actor is the user themself since they initiate the session. The transmission is activated either on the website or the app. There is no concrete recipient and the recipient type is either the platform or external friends. The recipient location is the website or app, where other people can view the live tracking session. The timing of the transmission is event-based because the live tracking session ends by itself after some time passed. There are no protection measures mentioned in the Garmin privacy* policy. The data processing is not commissioned and there is no intention to transmit the data to another country. (For the concrete privacy policy excerpt see Transmitting Data in Table 2.)

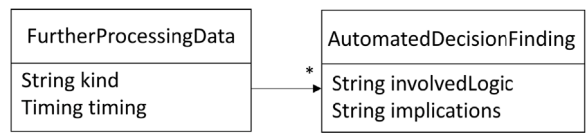

**Fig. 9.** Specialization of data processing: Further Processing Data (part of the conceptual model of privacy policies).

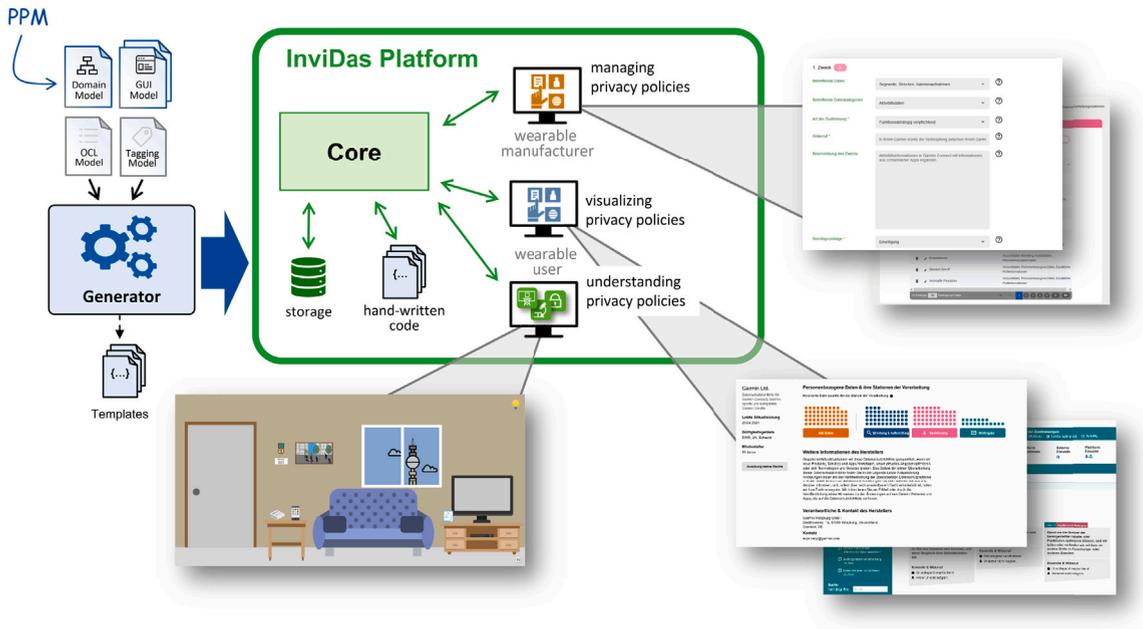

**Fig. 10.** MDSE process for creating an application to visualize privacy policies. The PPM acts as the domain model which is used as input for the generator.

*5.5.5. Further processing of data*

For further forms of processing data, the conceptual model (see Fig. 9) uses a joint concept `FurtherProcessingData`. An attribute describes the *kind* of the data processing (e.g., analyze or organize). A further attribute describes the *timing* when the data processing happens as with data transmission.

For some forms of data processing, the GDPR requires information about *automated decision-making* (including meaningful information about the *logic involved*, significance, and consequences of such processing). To prepare for this, the concept of further processing of data has dedicated attributes. If a privacy policy indicates more information about this form of data processing, it has to be captured in the scenario and the description text that each data processing concept possesses.

**Example.** An example of further data processing in the context of smartwatches happens in the scenario of calculating the calorie consumption of a user after an activity while they were using their smartwatch. All the necessary information (heartbeat, weight, height, sex of the user, etc.) to calculate the estimation was collected beforehand, but it has to be processed for the calculation. The actor, in this case, is the watch's user, because they initiate such a calculation. It is not an automated decision finding. (For the concrete privacy policy excerpt see Further Processing Data in Table 2.)

## 6. An example for generating applications using the privacy policy model

We use the developed conceptual model as input for the MontiGem code generator to create an application for the explanation and ergonomic visualization of privacy policies (see Fig. 10). The developed InviDas application includes three main parts: Input forms to manage privacy policies, explanations and ergonomic visualization of the privacy policy information, and an Escape game to make the potential challenges and results of data sharing more tangible (see Fig. 10, right). Within this article, we focus on the conceptual model and data needed to provide the necessary information for the first two parts. Using input forms, smartwatch vendors can add their privacy policy to the platform. This privacy policy information is then ergonomically explained and visualized for smartwatch users. The platform allows current and potential users of smartwatches to get to know the privacy policies of the devices in the form of interactive visualizations that are built on the respective instances of the conceptual model.

**Fig. 11.** Prototype screenshots of our web platform that allows smartwatch stakeholders to store data privacy information in the form of our conceptual model (text translated from German to English). (a) General information about the privacy statement, such as name or data controller. (b) Form to add a new data entry to the privacy statement and assign a corresponding data category if applicable. (c) Relevant details for the type of data processing.

The usage of our conceptual model provides the crucial foundation for a more flexible handling of privacy policies by storing the privacy information in a standardized and machine-readable format that can be used on a larger scale. While this reduces the effort of managing rigid textual privacy policies, the given structure in our model also supports stakeholders' compliance with the GDPR's requirements, mitigating existing confusion regarding the required disclosure of privacy practices [2].

To illustrate the usage of our conceptual model in the InviDas MDSE approach and its benefits for smartwatch stakeholders, we designed an input form (see Fig. 11) demonstrating an exemplary use case scenario of entering and storing privacy information based on our model's structure. The input form first requests basic details of a privacy policy (see Fig. 11(a)) including its name, the name of the data controller, a URL referring to the textual privacy policy, a valid from date and the minimum age for consenting to the privacy policy and using the smartwatch. Collected and processed data can be entered in and saved in Fig. 11(b). If applicable, the data entry can be assigned to a corresponding data category. For example, data collected during a specific fitness activity such as running (distance, pace, etc.) can then be grouped together by a "running activity data"-category. Specifics about the data handling are inquired in Fig. 11(c). First, the form of data processing has to be stated, followed by the person or party who collects the data, the technical processing form, the application use case scenario in which the user data is collected, the timing of the processing, and the place of processing. All input fields are accompanied by tooltips (displayed when hovering over the question mark icon) to provide additional explanations and assist stakeholders with entering the data. Mandatory information is indicated by the common practice of displaying an asterisk.

The InviDas application enables the ergonomic interaction of end users with privacy policy information. Fig. 12 shows a screenshot of the InviDas platform developed with the conceptual model. Concepts from the model such as different kinds of data processing, forms of agreement, or actors are utilized to visualize, categorize, filter, and highlight the information that is otherwise only available as natural text.

Building upon our model provides the necessary versatility for smartwatch manufacturers, designers, and developers for complying with the GDPR in conveying privacy information that is *"concise, easily accessible and easy to understand, and that clear and plain language and, additionally, where appropriate, visualization be used"*. As interactive and user-centered privacy statements are a crucial step towards full data sovereignty, our model plays a strong role in supporting current research in making privacy information more accessible [79–81] and more adaptable to the needs of the heterogeneous end-user group of smartwatches, including citizens of different ages and various purposes of use [82–84].

## 7. Discussion

Our experiences within the modeling and visualization process as well as discussions with wearable manufacturers and lawyers led to several observations discussed in the following.

### 7.1. Mapping of our conceptual model to existing privacy policy taxonomies

Our conceptual model supports the majority of concepts of the taxonomy used by Morel and Pardo [60] which is based on the taxonomy provided by Wilson et al. [61]. In more detail: The taxonomy items "first party collection" and "third party collection" from Morel and Pardo [60] align with our concept *CollectingData* (see Section 5.5.1). The distinction between first and third party is captured in the *actor* attribute (see Section 5.4). The actor Manufacturer can be used for first party and the actor External Data Recipient can be used for third party collection (see Section 5.4). The "Legal Basis" taxonomy item is represented in our model by the concepts *LawfulBasis* and *LawfulBasisType* (see Section 5.4). The taxonomy item "Data Subject Rights" is captured in our conceptual model with the concepts *Right* and *RightType* (see Section 5.2). The taxonomy item "Data Retention" aligns with our concept of *StoringData* (see Section 5.5.2). The concepts *StoringData* and *TransmittingData* both have an attribute regarding data security which is another item of Morel and Pardo's taxonomy (see Section 5.5.4). Lastly, the taxonomy item "Policy Change" is captured in our conceptual model by the attribute *updatePolicies* of the *PrivacyPolicy* concept (see Section 7.3).

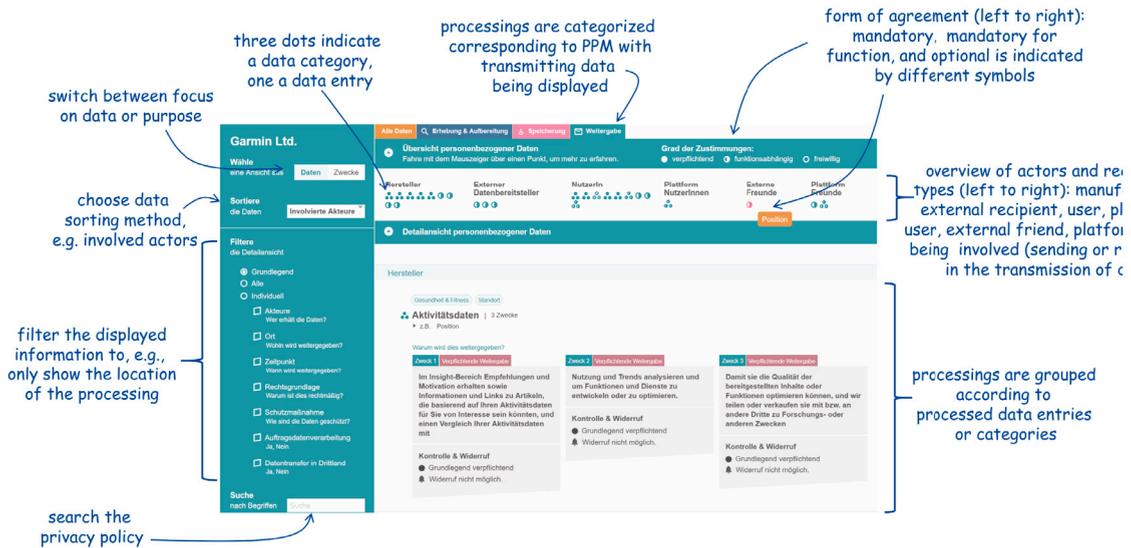

**Fig. 12.** A screenshot of our web platform that allows smartwatch users to ergonomically interact with stored data privacy information. The conceptual model enables different ways to categorize, filter, and highlight the data. The platform (only available in German) can be accessed at https://datenschutz-lotsin.de/home.

We partly covered the "Specific Audiences" aspect mentioned in Wilson et al. [61], namely by using *minimum age* and an attached *region* as parts of the *PrivacyPolicy* concept (see Section 7.3). They mention "practices that pertain only to a specific group of users" - as specific groups only the age restriction and geographic region are relevant for smartwatches. Wilson et al. [61] also suggest "do not track" as a part of their taxonomy. The "do not track" HTTP header field is a generic, technical solution to deny tracking when someone makes an HTTP request. This field is relevant when the user is given control over the request, e.g., via browser settings. In the smartwatch context, it is less relevant because usually the smartwatch or a specific app handles all the communication to the server and the manufacturer is in control of it. Because of that, there is no need for a generic solution and usually, the user is not given control over such a technical detail in the communication. Morel and Pardo subsume these aspects in their analysis as "other" concepts, "as they are occasional items" [60].

Many concepts our conceptual model includes explicitly, such as the *DataProtectionOfficer*, the *Controller* of a region, and their *Contact* information Section 5.2, are only loosely represented under the taxonomy item "Other" from Wilson et al. [61].

As our conceptual model is aimed to be used within an MDSE approach, we require a greater level of detail of the used concepts and their relationships than provided in existing taxonomies. This enables us to cover the detailed descriptions in concrete privacy policy instances provided by legal experts.

### 7.2. Findings from creating instances of the PPM

To create instances of the PPM, e.g., for the InviDas project, we read the specific privacy policies and followed a procedure, which is described below. During the procedure, we faced challenges, that we think are not limited to our approach but are problems that apply to a broader scope. First, you start with the creation of the privacy policy object, because all other objects are transitively linked to it. Then, you proceed with the region(s) and the objects attached to it, i.e., how customers can exercise their rights and the data protection officer. This information is typically relatively easy to find since the information and its structure are enforced by laws such as the GDPR and since they are similar across different manufacturers. At last, you have to perform the most difficult task creating all data processing objects, their purposes, and the data attached to them. In the following, we describe our challenges with this task.

One big challenge for fitting the privacy policies of the various manufacturers into instances of the PPM was that they all differed in structure. Garmin's privacy policy, for example, first describes the scenario in which the data processing occurs. In addition, it lists the data entries and categories involved and it breaks them down into a list of purposes including their form of agreement, and their legal basis. The structure of FitBit's privacy policy on the other hand includes three sections concerning the description of FitBit's data processing. The first section is for data collection, the second one is for data usage, and the third one is for data sharing. This is also a concise and valid way to structure the privacy policy, but it makes it harder to connect the processed data with a purpose because collecting data always has a purpose that includes the data's usage. So to relate the collected data with its purpose, we had to go back and forth between the two sections.

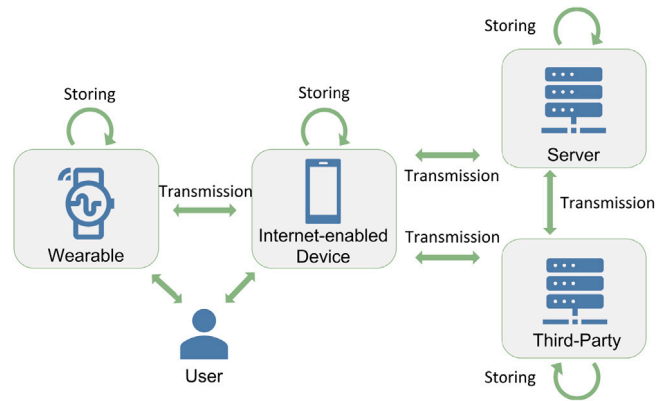

**Fig. 13.** Coming from a technical background we initially expected the information about data processing contained in privacy policies to be more technically detailed giving users insights about where exactly their data is processed (adapted from [85]).

Another challenge was specific to us fitting the privacy policies into their PPM instances. We could only use the information available from the public privacy policies. An employee of the company could get more insights into technical specifics, which our model can capture. This includes, for example, the protection measures for data transmission. Adding this information could help tech-savvy users in choosing which smartwatch to buy and which services to use from a specific manufacturer.

Oftentimes, the information we expected to be explicitly mentioned in the privacy policies was not explicitly stated, e.g. revocation options. It is required by law to state possible options for revoking one's consent to specific data processing, but privacy policies do not contain this information for each data processing they describe. We concluded that the only revocation option was to stop using the wearable or service and request deletion of your data at the manufacturer.

Finally, we initially expected a privacy policy to contain detailed information about the data processing of a specific smartwatch, e.g., from the smartwatch to an internet-enabled device, to a server, or to third-party applications (see Fig. 13 adn [85] ). We assumed, that we could map the information in privacy policies to these different steps of data processing. The reality deviated from our expectations, in various regards. Privacy policies only contain information about legal data transmissions and not technical ones (see Section 5), e.g., the transmission from a smartwatch to an internet-enabled device such as your smartphone. Furthermore, privacy policies hardly contain information about how and where data is stored. In Garmin's privacy policy, e.g., is a single paragraph about the storing of all *personal* data. In this respect, privacy policies do not meet the expectations of device users.

### 7.3. General findings and limitations

*Discussing our requirements to the PPM.* In Section 3, we introduced the requirements we took into account while creating the PPM. Requirement R1 was trivially satisfied by choosing a data structure modeling language such as the UML Class Diagram language along with common data types like boolean, String, Date, etc. To ensure that requirement R2 is satisfied, we did a thorough analysis of the current law in the European Union regarding privacy policies and legal experts checked our conceptual model. To satisfy requirement R3, we analyzed examples of privacy policies of 7 different, well-known smartwatch manufacturers (Section 4). To further validate requirement R3, we took the privacy policies of Garmin and Fitbit and successfully transformed both into an instance of the PPM (see object diagrams at [78]).

*Automated validation of instances of the PPM.* The PPM currently allows its instances to be inconsistent with current law in regions like the European Union. If data is transmitted to a country that is not part of the European Union and there is no adequacy decision for that country, the GDPR requires safeguards for this country [1, (cf. Art. 13 §1.f and Art. 46 §1)]. OCL could be used to ensure that such a safeguard is captured in the PPM instance. Listing 1 shows how the requirement can be modeled as invariant in OCL.

```
context TransmittingData tD inv:
  forall region in tD.recipientRegions
    (region instanceof OtherCountryRecipients) implies
      tD.safeguardsForRegions.isPresent()
```

**Listing 1** OCL invariant that ensures, each TransmittingData which transmits data into another Region has a safeguard.

Such validation helps to ensure the correctness of the transformation from privacy policy to PPM instance. Considering privacy policies contain descriptions of compliance with various laws, they may contain inconsistencies and ambiguity. Because of this, OCL on its own is not sufficient to also validate a privacy policy based on its PPM instance. For this task more complex requirement languages are necessary, allowing for the modeling of uncertainty and relaxation of requirements, such as the language RELAX [86].

*Raising awareness on the manufacturer side.* In addition to the preparatory functionalities for technical implementation, the conceptual model presented has the power to convey processes and interrelationships of data protection and raise awareness of the topic of data protection at the manufacturer and on the user side. By summarizing the more complex processes, parameters, and interrelationships that are necessary for gaining knowledge, the model reduces data protection complexity to the crucial content. This intellectual effort no longer has to be performed by the recipient, which facilitates information access. However, easy access to information also requires the linking of visual properties with the model structure, which is familiar not only to software experts but also to elderly people and those who are not familiar with technology.

*Limited attributes for privacy choices.* While our conceptual model covers the extensive corpus of information of current textual privacy policies, it also reflects the existing limited availability of privacy choices. Based on our analysis, the conceptual model only contains privacy choices connected to legal purposes, which are described by the attributes *form of agreement* and *revocation* of an agreement. To further extend the control options and enable interactive visualization to offer end-users detailed control over their data privacy, additional attributes need to be identified in future user-centered research. For instance, privacy choices could be extended by a period attribute to which the data collection or processing can be limited.

*Flexibility for data entries and categories.* The analysis of the existing privacy policies from different smartwatch manufacturers has shown that there is no common practice for putting specific data entries in the same data category or to name the data categories the same, e.g., activity data might include geoinformation, activity types, the activity degree, or active minutes (or not all of them — it is often underspecified which attributes are explicitly included) where others put geoinformation in a geodata category or active minutes to device information. This makes the visualization more challenging, especially when comparing different manufacturers.

*Completeness.* We do not guarantee that all information of a privacy policy can be manifested in the model, as legal texts are still the only legally approved way to define legally binding texts. In our opinion, legal informatics should drive the idea that models should also be recognized as legally binding texts as suggested e.g., in [87], however, to realize this is probably a long way off.

*Extension of the MDSE approach.* Using the conceptual model for privacy policies is a first step towards considering privacy directly within the MDSE process. While this approach provides great support for automating the usage of the correct concepts within the database and the transport layer of the application, it would be worth investigating MDSE approaches beyond that, e.g., ensuring with model-driven methods that a generated application adheres to the statements in the privacy policy, or validating that it does so. This requires the cooperation of experts in data, software, and process modeling [88] with those in software architectures, privacy, and security. In addition, the automated generation of ergonomically improved user interfaces to explain privacy policies would reduce the amount of hand-written additions needed in the InviDas platform. However, this would require the cooperation of experts from MDSE with human–computer interaction research.

### 7.4. Threats to validity

We are discussing internal threats to validity, i.e., whether our method to create and validate the PPM makes a difference or not, whether there is sufficient evidence to support our claim, and external threats to validity, i.e., the generalizability of the outcomes.

*Validation of the model with experts.* The present model was developed based on content analyses of existing privacy statements of well-known smartwatch manufacturers. The validity or invalidity of this model additionally needs to be verified, i.e., evidence of its correctness needs to be provided. The preceding work already includes an initial expert interview with one manufacturer and one experienced legal expert. However, to reduce the risk of overlooking relevant model parameters, making application errors such as the unsuitable application of a model to a system, and disregarding areas of validity of the model, further expert validations with larger samples are required.

*Evaluation of the usefulness of the model for MDSE processes.* To show how useful the conceptual model is for developers in MDSE processes, we have used it to create the platform. However, evaluating the usefulness of the conceptual model in various use cases and development projects poses challenges: developers have diverse backgrounds, and each project has different requirements. The different perspectives and expertise brought by these developers as well as the project requirements may lead to varying evaluations and results. Further evaluations would require corresponding funding of projects and further examples from practice, which, however, are not currently available. Since our conceptual model is publicly available, this allows for more practical examples to be evaluated in the future.

*Generalizability.* Up to now, we have applied the conceptual model only in the application domain of smartwatches. However, most concepts apply to different domains, e.g., wearables in general or other data-collecting devices including IoT devices. Some privacy policies, e.g., from Garmin and Fitbit, do not differentiate between data from smart wearables and smartwatches, as they have one privacy policy for all their smart wearable products. While the InviDas project focuses on smartwatches, we can assume that the conceptual model is applicable to all smart wearable products. What is smartwatch specific in the conceptual model are enumerations that might need to be adapted for different data-collecting devices.

### 7.5. Alternative methods to make privacy policies better understandable

Using a conceptual model and MDSE approach is only one possible way how to make information from privacy policies better understandable. Traditional natural language processing as well as Large Language Models provide further alternatives.

*Using natural language processing.* Natural language processing tools may be able to simplify the language of privacy policies or analyze them for certain legal terms. In a further step, their output has to be structured and somehow formalized to be able to be used for a consistent and comparable display of information of different privacy policies. Thus, natural language processing tools cannot replace the PPM but could instead support the transformation of textual privacy policies to PPM instances.

*Using Large Language Models (LLMs).* As an alternative approach, one could use LLMs to simplify the language of privacy policies for end users or to analyze and compare different privacy policies. From a legal point of view, using LLMs contains risks that are not acceptable for legally binding texts, e.g., non-determinism of results or a lack of correctness of the output [89,90]. However, LLMs could be used to support the process of transforming textual privacy policies to PPM instances but still require the check and validation of legal experts in a subsequent step. In addition, LLMs could be used to generate web applications from natural language [91]. An extension to applications that use privacy-relevant data would be worthwhile to investigate.

## 8. Conclusion

Within this paper, we have introduced an approach to support the building of applications that improve the understandability of privacy policies and data processing of smartwatches. We introduce a conceptual model for privacy policies for smartwatches and show its use to generate an application in a model-driven software engineering approach. We have defined the concepts in the conceptual model based on an analysis of the privacy policies of seven smartwatch manufacturers. These concepts in the model were discussed with lawyers for their correctness. In the next step, we have instantiated the conceptual model with concrete data to show its applicability. The textual version of the conceptual model is then used in a MDSE approach to create the InviDas platform for further explanation and ergonomic visualization of related concepts.

Other developers can use the conceptual model to handle information in privacy policies in a structured way in their development processes. Reusing the model allows them to easier add privacy policy-related concepts in the data structure of applications in the long run. In addition, the complete visualization of data protection relationships requires an extensive collection of correspondingly valid data and context information from different domains. The presented data model as part of future systems could serve as the framework for the collection of data protection meta-information. However, its validity for other contexts would then have to be demonstrated.

Finally, when the goal is to promote an understanding of privacy policies to end users as in the InviDas platform in a fully automated manner, there is a long way to go. There is further research in human–computer interaction and MDSE needed on how to visualize complex, legal information for end users in such a way that all types of citizens with different abilities and backgrounds can make informed decisions.

## CRediT authorship contribution statement

**Constantin Buschhaus:** Writing – review & editing, Writing – original draft, Visualization, Validation, Software, Resources, Methodology, Data curation, Conceptualization. **Arvid Butting:** Writing – review & editing, Writing – original draft, Visualization, Validation, Software, Resources, Methodology, Data curation, Conceptualization. **Judith Michael:** Writing – review & editing, Writing – original draft, Visualization, Validation, Supervision, Software, Resources, Project administration, Methodology, Funding acquisition, Data curation, Conceptualization. **Verena Nitsch:** Writing – review & editing, Validation, Supervision, Resources, Project administration, Methodology, Funding acquisition, Conceptualization. **Sebastian Pütz:** Writing – review & editing, Writing – original draft, Visualization, Validation, Software, Resources, Methodology, Data curation, Conceptualization. **Bernhard Rumpe:** Writing – review & editing, Validation, Supervision, Software, Resources, Project administration, Methodology, Funding acquisition, Conceptualization. **Carolin Stellmacher:** Writing – review & editing, Writing – original draft, Visualization, Validation, Software, Resources, Project administration, Methodology, Data curation, Conceptualization. **Sabine Theis:** Writing – review & editing, Writing – original draft, Visualization, Validation, Supervision, Software, Resources, Project administration, Methodology, Funding acquisition, Data curation, Conceptualization.

## Declaration of competing interest

The authors declare that they have no known competing financial interests or personal relationships that could have appeared to influence the work reported in this paper.

## Acknowledgments

This article describes research first identified as part of the research project "InviDas – Interaktive Datenvisualisierungen für eine digitalrechtliche Entscheidungsfindung" (grant number: 16SV8536) funded by the German Federal Ministry of Education and Research (BMBF), Germany and supervised by VDI/VDE Innovation + Technik GmbH.

## Appendix

See Table 2.

**Table 2**

The quotations in column 2 are taken from a Garmin privacy policy of 2022 on which the examples in Section 5 were based. Different excerpts are grouped according to their corresponding example. Each bullet point is an excerpt from a different passage of the privacy policy. The whole privacy policy including highlighted excerpts, is at Zenodo [78].

| Concept of the PPM | Relevant text excerpts taken from the Garmin privacy policy |
| --- | --- |
| Privacy Policy | – "Privacy Policy for Garmin Connect and Compatible Garmin Devices Last Updated: August 31, 2022"<br>– "Any changes will become effective upon our posting of the revised Privacy Policy."<br>– "Recent Policy Versions April 29, 2021 June 5, 2020" |
| Region<br>Data Controller<br>Data Protection Officer | – "Data Controller and Data Protection Officer If you reside in a country in the EEA, U.K., or Switzerland, then your personal data collected by Garmin is controlled by Garmin Würzburg GmbH, Beethovenstr. 1a, 97080 Würzburg, Germany. The company's EU Data Protection Officer can be reached by email at euprivacy@garmin.com."<br>– "If you reside in the EEA, U.K., or Switzerland, you have the right, subject to the conditions set out in the General Data Protection Regulation ("GDPR") or other applicable law, to request from Garmin access to and rectification or erasure of your personal data, data portability, restriction of processing of your personal data, the right to object to processing of your personal data, and the right to lodge a complaint with a supervisory authority. For more information about these rights, please visit the European Commission's "My Rights" page relating to GDPR, which can be displayed in a number of languages. If you reside outside of the EEA, U.K., and Switzerland, you may have similar rights under your local laws." |
| Rights | – "To request access to or rectification, portability, or erasure of your personal data, or to delete your Garmin account, please visit our Account Management Center." |
| General Data Processing<br>Purpose<br>Lawful Basis | – "We also process your email address to associate it with your Garmin account when you interact with our customer support representatives. The legal ground for this processing is our legitimate interest in providing quality customer support." |
| Data Entry<br>Data Category | – "We process your email address and password because you use your email address and password to sign in to your account."<br>– "We also process your email address for the purpose of sending you important information"<br>– …<br>– "Categories of Personal Data Processed by Garmin Personal data that is processed when you create a Garmin account" |
| Collecting Data | – "Personal data that is processed when you add a Garmin device to your Garmin account: When you add certain Garmin devices to your Garmin account, we ask for additional information, such as your gender, height, weight, birthdate, activity level (low, medium, or high), and normal bed and wake times." |
| Storing Data | – "Retention of Personal Data We will retain your personal data as long as your Garmin account is considered to be active or in accordance with applicable law and regulatory obligations. In addition, see below under "Your Rights" for a description of your right of erasure."<br>– "For customers outside of mainland China, when you create a Garmin account, add personal data in your account profile, or upload data to your Garmin account, your personal data will be collected and stored on servers in the U.S., U.K., and/or Australia." |
| Deleting Data | – "If we learn that we have collected personal data from a child under the age of 13 in the U.S. or under 16 in the rest of the world, we will take steps to delete the information as soon as possible." |
| Transmitting Data | – "YOUR REAL-TIME TRACKING INVITEES: Some Garmin devices include features, such as LiveTrack, that enable you to send a link to people of your choice that allows them to see the real-time location of your device. Because anyone with access to the link will be able to see the real-time location of your Garmin device, you should use caution in determining to whom you want to send the link and be sure that you trust them to not send the link to others whom you do not want to be able to view the location of your Garmin device."<br>**Author Remark:** Information about the timing could only be found in a source explaining LiveTrack https://www8.garmin.com/manuals/webhelp/drive51-61/EN-US/GUID-1ABE36FC-00D9-4206-AE90-D66E59C3F022.html |
| Further Data Processing | – "Personal data that is processed when you choose to upload or add your data to your Garmin account"<br>– "You can choose to upload from your device or, in some cases, manually add activities (e.g., runs, walks, bike rides, swims, hikes, gym activities, etc.) and activity data (e.g., steps, distance, pace, activity time, calories burned"<br>– "We process this data, if you choose to upload or add it to your Garmin account, to enable you to analyze this data" |

**Data availability**

The used data is published at Zenodo [78].

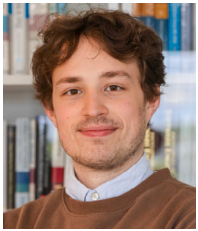

**Constantin Buschhaus** received his B. Sc. and M. Sc. degrees in computer science from the RWTH Aachen University, in 2020 and 2022. He is a research assistant and Ph.D. candidate at the Software Engineering chair at RWTH Aachen University since 2022. His research focuses on model-driven development, artifact models, and software project analysis based on artifact models.

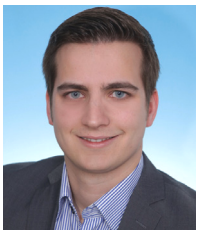

**Arvid Butting** received his B. Sc. and M. Sc. degrees in computer science from the RWTH Aachen University, in 2014 and 2016. His Ph.D. thesis at the Software Engineering chair at RWTH Aachen University was about the systematic composition of language components in the language workbench MontiCore. His research interests cover software language engineering, software architectures, and model-driven development.

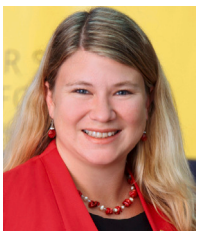

**Judith Michael** is PostDoc and team leader at the Software Engineering chair at RWTH Aachen University, Germany. Her Habilitation at RWTH Aachen gave insights into the model-driven engineering of digital twins with informative and assistive services. Her Ph.D. thesis at Universität Klagenfurt was about cognitive modeling for assistive systems. In her research, she is focusing on the engineering of complex, long-lasting, software-intensive systems within various application domains in an integrated approach with different disciplines. Her recent work deals with software architectures of assistive and information systems, digital twin engineering and digital shadows in production, privacy-preserving system design, and human behavior modeling.

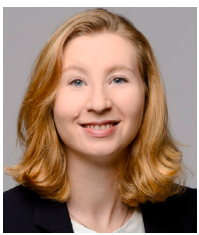

**Verena Nitsch** studied applied psychology at Charles Sturt University in Australia and the University of Central Lancashire in the UK before completing her master's degree in industrial and organizational psychology at Manchester Business School. She completed her doctorate in engineering in the field of human-technology interaction at the Bundeswehr University Munich, where she became a professor of Cognitive Ergonomics and headed the Human Factors Institute (IfA) from 2016 to 2018. Since June 2018, she has been full professor and Director of the Institute of Industrial Engineering and Ergonomics at RWTH Aachen University (IAW). She is also currently Head of the Department of Product and Process Ergonomics at the Fraunhofer Institute for Communication, Information Processing and Ergonomics FKIE.

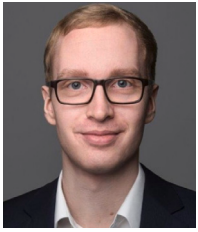
**Sebastian Pütz** received his bachelor's degree in psychology from RWTH Aachen University and his master's degree in human factors engineering and ergonomics from the Technical University of Munich. He is a research associate and Ph.D. candidate at the Department of Ergonomics and Human-Machine Systems at the Institute of Industrial Engineering and Ergonomics (IAW) at RWTH Aachen University. Sebastian's research focuses on the cognitive ergonomics of human-machine interfaces with the goal of optimizing operator workload in sociotechnical production systems.

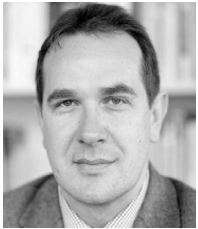
**Bernhard Rumpe** is heading the Software Engineering chair at RWTH Aachen University, Germany. His main interests are rigorous and practical software and system development methods based on adequate modeling techniques. This includes agile development methods as well as model engineering based on UML/SysML-like notations and domain specific languages. He also helps to apply modeling, e.g. to autonomous cars, human brain simulation, BIM energy management, juristic contract digitalization, production automation, cloud, and many more. He is author and editor of 34 books including "Agile Modeling with the UML" and "Engineering Modeling Languages: Turning Domain Knowledge into Tools".

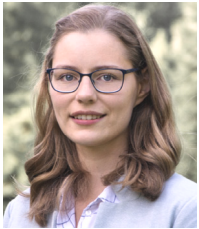
**Carolin Stellmacher** earned her bachelor's degree in computer science from the Hamburg University of Applied Sciences and her master's degree in computer science and human–computer interaction from the University of Bremen. She is presently at the University of Bremen pursuing a Ph.D. in human–computer interaction and holding a position as a research associate. Part of her research focuses on developing interactive technologies that effectively communicate complex data spaces from fitness and health applications to users and foster motivation for managing their own data privacy.

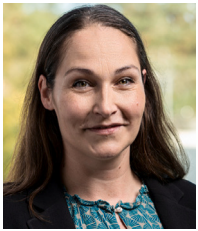
**Sabine Theis** is the lead of the Human Factors in Software Engineering research group at the German Aerospace Center (DLR). At a higher level, her work focuses on evaluating and characterizing data and information visualization systems and techniques from a human factors perspective. Currently, she investigates the potential of provenance visualizations for the trustworthiness and explainability of artificial intelligence in air traffic control. Past projects include InviDas, investigating the potential of interactive data visualizations to support individual digital sovereignty of users of AI-infused wearables and the PCompanion project on user-centered development of a Parkinson's Disease monitoring and management application, including interactive data visualizations.